\documentclass{article}
\usepackage{amsmath,epsfig}
\usepackage[preprint]{spconfa4}
\usepackage{multirow}
\usepackage[justification=centering]{caption}

\let\OLDthebibliography\thebibliography
\renewcommand\thebibliography[1]{
  \OLDthebibliography{#1}
  \setlength{\parskip}{0pt}
  \setlength{\itemsep}{0pt plus 0.3ex}
}

\begin{document}\sloppy

% Example definitions.
% --------------------
\def\x{{\mathbf x}}
\def\L{{\cal L}}

% Title.
% ------
\title{Residual Attention Based Network for \\Automatic Classification of Phonation Modes}

%
% Address.
% ---------------
\name{Xiaoheng Sun$^{\ast}$, Yiliang Jiang$^{\ast}$, Wei Li$^{\ast \dagger}$\thanks{This work was supported by National Key R\&D Program of China (2019YFC1711800), NSFC (61671156).}}

\address{
$^{\ast}$School of Computer Science and Technology, Fudan University, Shanghai, China; \\
$^{\dagger}$Shanghai Key Laboratory of Intelligent Information Processing, Fudan University, Shanghai, China.
}

\maketitle

\begin{abstract}
Phonation mode is an essential characteristic of singing style as well as an important expression of performance. It can be classified into four categories, called neutral, breathy, pressed and flow. Previous studies used voice quality features and feature engineering for classification. While deep learning has achieved significant progress in other fields of music information retrieval (MIR), there are few attempts in the classification of phonation modes. In this study, a Residual Attention based network is proposed for automatic classification of phonation modes. The network consists of a convolutional network performing feature processing and a soft mask branch enabling the network focus on a specific area. In comparison experiments, the models with proposed network achieve better results in three of the four datasets than previous works, among which the highest classification accuracy is 94.58\%, 2.29\% higher than the baseline.
\end{abstract}
\section{Introduction}
\label{sec:intro}

There are various instruments in the world, among which the human voice is the most amazing and prevalent. From the objective point of view, different vocal fold conditions or shapes produce different vocal production qualities, also known as phonation modes. 

According to Johan Sundberg \cite{1.1Sundberg}, there are four phonation modes in singing: breathy, neutral, flow and pressed. In breathy phonation, there is a reduced vocal fold adduction and minimal vocal fold contact area, which contribute to higher level of turbulent noise and higher harmonic-to-noise ratio (HNR) \cite{1.1Sundberg}. In neutral phonation, the closed phase is somewhat shortened and the airflow during the opening phase is considerably increased \cite{1.1Childers}. Pressed phonation displays a long closed phase, with reduced airflow during the opening phase \cite{3.1prout}. Flow phonation is produced by lower larynx, where the maximal airflow is achieved retaining a closure of the vocal folds during the closed phase \cite{1.1Sundberg}. Unlike other phonations, it is generally considered as a singing skill.

Automatic classification of phonation modes is of great significance. On one hand, the use of different phonation modes indicates different control capability over glottis and vocal folds. E.g., the vocal music teachers can judge the students' singing level from phonation modes; analysis of irregular phonation modes can help to diagnose certain pathological voice. On the other hand, detection of different phonation modes can be the basis of other high-level music information retrieval (MIR) tasks, such as singing evaluation, music emotion recognition, musical genre classification, etc.

In this paper, we propose a novel deep learning method to automatically classify the phonation modes. A Residual Attention based convolutional neural network is built to automatically extract discriminative features from Mel-spectrogram. Gradient-weighted class activation map (Grad-CAM) is utilized for analyzing which part of the spectrogram is important for the classification.

The rest of the paper is organized as follows. Section 2 introduces the related work. Section 3 describes the methodology in detail. In Section 4, experimental results are presented and analysed, and in Section 5, we make further conclusion.

\section{Related work}

Several studies have investigated the classification of phonation modes from the perspectives of aerodynamics, voice quality and spectrum.

The aerodynamic features reflect the principle of pronunciation and usually collected by aerodynamic detector. For example, the glottal resistance is the ratio of the difference between the glottal pressure and the average glottal airflow, which can reflect the pressure under the glottis and the area of the glottis. The vocal efficiency is the ratio of the sound intensity to the average glottal airflow rate, which is determined by factors such as the vocal chord function, the amplitude of the vocal cord vibration, and the uniformity of the pressure in the larynx. Grillo and Katherine proved the effectiveness of laryngeal resistance and vocal efficiency in distinguishing different phonation modes \cite{1.1grillo}. Nonetheless, collecting aerodynamic characteristics is a complicated and expensive task.

Voice quality features are usually calculated from inverse filtering. In the speech vocalization scenario, \cite{1.1millgaard} proved that the acoustic parameters such as normalized amplitude quotient (NAQ), amplitude, glottal closed entropy, energy ratio around 1000 Hz, etc. have a certain consistency with the results of expert judgment in discriminating phonation modes. \cite{1.1airas} \cite{1.1alku} showed that compared with the traditional glottal amplitude entropy feature, the normalized amplitude quotient, which characterizes the glottal excitation, can distinguish the four phonation modes more robustly. Standardized amplitude quotients achieved a 73\% consistency score with expert judgement on singing vocal pressure values \cite{1.1sundberg2}. Acoustic features such as peak slope \cite{1.1kane1}, maximum dispersion quotient (MDQ) \cite{1.1kane2}, and significant cepstral peaks can be used as remarkable features to differentiate breathy and pressed phonation \cite{1.1hille}. Because of inverse filtering problems from singing voice, voice quality features alone are not sufficient for classification \cite{1.1kadiri1}I. In order to improve the time-frequency resolution, Sudarsana used the improved zero-frequency filtering method and the cepstrum coefficient, proposing a zero-time window method \cite{1.1kadiri2}.

In recent studies, researchers have focused more on spectral features. For example, harmonic amplitudes, formant frequencies bandwidths and amplitudes, harmonic-to-noise ratio, etc., are combined with sound quality characteristics to classify the phonation mode \cite{3.1rouas}. The frequency domain features can perform well under certain scenarios, but due to different vowel, pitch and other conditions, some features may not applicable in all situations.

\section{METHODOLOGY}

In this section, we first describe the dataset and data processing methods. Then we introduce the detailed design of network architecture.

\subsection{Dataset}

To compare with the work in \cite{3.1Yes}, we use the same four datasets and divide them into training and test sets in the same way. Each dataset contains recordings of individual sustained vowels with different pitches and phonation modes.

\textbf{Dataset-1} \cite{3.1prout}: The first available dataset (DS-1) for phonation mode is published by Proutskova in 2013. The dataset contains a total of 909 audio clips, which were sung by a professional Russian soprano singer. The recording includes 9 Russian vowels, ranging in pitch from $A3$ to $C6$. 
	
\textbf{Dataset-2} \cite{3.1rouas}: The second phonation dataset (DS-2) is published by Rouas and Ioannidis in 2016. The dataset contains 487 recordings sung by a professional baritone singer. The pitch varies from ${\settoheight{\dimen0}{C}C\kern-.05em \resizebox{!}{\dimen0}{\raisebox{\depth}{$\sharp$}}}2$ to ${\settoheight{\dimen0}{G}G\kern-.05em \resizebox{!}{\dimen0}{\raisebox{\depth}{$\sharp$}}}4$.

\textbf{Dataset-3} \cite{3.1Yes}: The third dataset (DS-3) used for automatic phonation classification is recorded by Universitat Pompeu Fabra (UPF). It includes 515 recordings sung by a professional female soprano singer in 2018. The pitch varies from $A3$ to $C6$.

\textbf{Dataset-4} \cite{3.1Yes}: The fourth dataset (DS-4) is also published by UPF. There are 240 recordings sung by a professional female soprano singer in this dataset. The pitch varies from ${\settoheight{\dimen0}{F}F\kern-.05em \resizebox{!}{\dimen0}{\raisebox{\depth}{$\sharp$}}}3$ to $F4$.

\begin{table}[t]
\begin{center}
\caption{The number of audio clips in the four datasets. After data augmentation, the training data size has increased by about twice, while the test data size remains the same.} \label{tab:table1}
\setlength{\tabcolsep}{1.6mm}{
\begin{tabular}{|c|c|c|c|}
  \hline
   & Audio data type & Original data & Augmented data
  \\
  \hline
    \multirow{2}{*}{DS-1} 
      & Training & 727 & 1563 \\
	~ & Test & 182 & 182 \\
    \hline
	\multirow{2}{*}{DS-2} 
      & Training & 389 & 877 \\
	~ & Test & 98 & 98 \\
	\hline
	\multirow{2}{*}{DS-3} 
      & Training & 412 & 925 \\
	~ & Test & 103 & 103 \\
	\hline
	\multirow{2}{*}{DS-4} 
      & Training & 192 & 548 \\
	~ & Test & 48 & 48 \\
 
  \hline
\end{tabular}
}
\end{center}

\end{table}

\begin{figure}[t]
\begin{minipage}[b]{1.0\linewidth}
  \centering
  \centerline{\epsfig{figure=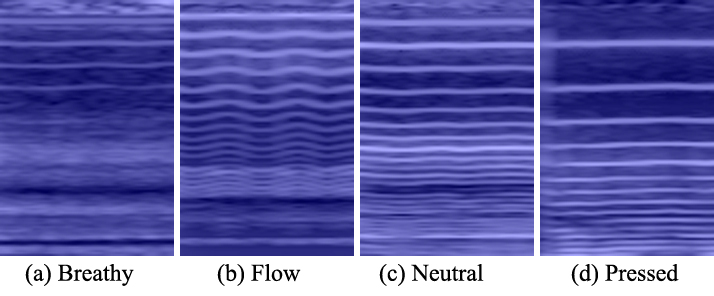,width=8.5cm}}
  \vspace{-0.1cm}
  \caption{The Mel-spectrograms for different phonation modes. Frequencies are shown increasing up the vertical axis, and time on the horizontal axis. The brightness increases with the magnitude.}
  \label{fig:fig_mel_spectrogram}
\end{minipage}
\end{figure}

\begin{figure*}[t]
\begin{minipage}[b]{1.0\linewidth}
  \centering
  \centerline{\epsfig{figure=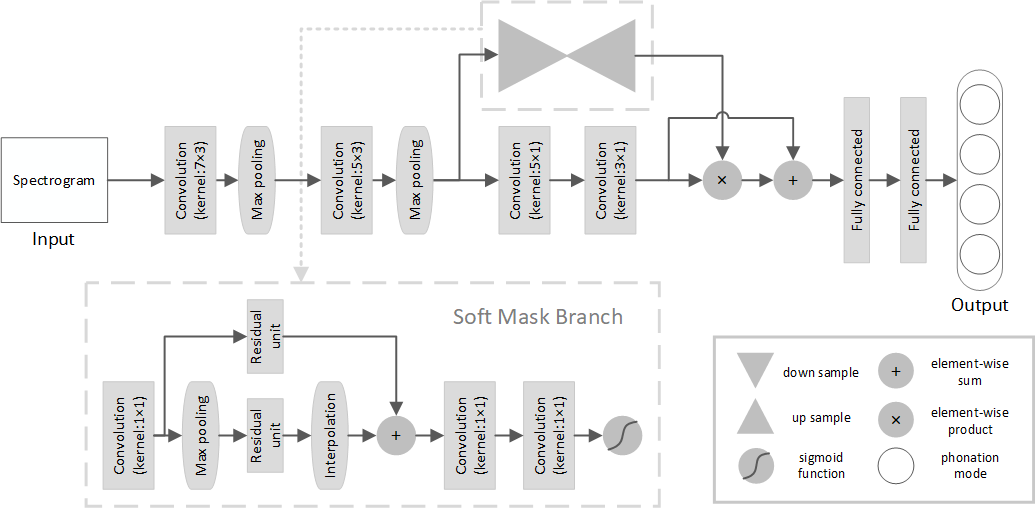,width=16cm}}
  \vspace{0.3cm}
  \caption{The illustration of the proposed Residual Attention based network with soft mask branch.}
  \label{fig:fig_network}
\end{minipage}
\end{figure*}

\subsection{Data preprocessing and augmentation}

First, the original audio data is resampled to 44.1kHz and blank audio segments are removed in the preprocessing. Then, because the Mel-scaled frequency match closely the human auditory perception, we choose Mel-spectrograms as raw feature maps. To extract feature maps, we use 12.5\%-overlapping windows of 2048 frames, and transform each window into a 128 band Mel-scaled magnitude spectrum. The Mel-spectrograms of four phonation modes are shown in Figure \ref{fig:fig_mel_spectrogram}.

To increase the amount of data available for training, we perform data augmentation on training sets. In \cite{3.1Yes}, only the middle 500 ms segments of audios are selected as valid data, and the rest are discarded because they hold that the phonation mode is not stable on these segments. But in fact, this approach is a waste of useful data to some extent. In contrast, we only remove the potentially unstable 128 ms at the head and tail of each spectrogram, and segment the spectrogram with a 500 ms window and 128 ms overlapping length. By this way, the training data is expanded about twice, as shown in Table \ref{tab:table1}. Note that none of the data contains blank segments, and there are no augmented data in test sets.

\subsection{Frequency-biased filters}

In the implementation of convolutional neural network (CNN), squared filters are most commonly used as the convolution kernel. Squared filters are easy to understand and can be widely used in various scenarios. Research in computer vision has achieved significant results by using CNN with squared filters, but adjusting the shape of filters in different tasks is still an effective tuning method.

\cite{2.3Pons} explores the application of different filter shapes in MIR research, and points out that in the processing of audio, using \textit{1-by-n} or \textit{n-by-1} filters can sometimes achieve better results. Unlike the images in the field of computer vision, the two dimensions of audio spectrograms have different meanings, which respectively represent the changes in the frequency and time domain of the sound. \textit{1-by-n} filters, also known as temporal filters, are more conducive to learning high-level features in the time domain, while \textit{n-by-1} filters, also known as frequency filters, are more conducive to learning high-level features in the frequency domain.

In phonation mode classification, by observing the spectrogram, it can be found that different phonation modes show different energy distribution in frequency bands, and the difference in the frequency domain is more significant than in the time domain. Therefore, this experiment adopts the idea mentioned above, using \textit{n-by-m (m $<$ n)} filters, which we call frequency-biased filters, to make the model more focus on the learning of frequency-domain features.

\subsection{Residual Attention based networks}

\begin{table*}[t]
\begin{center}
\caption{Average F-measure values of the proposed methods
and other designed experiments \protect\\in comparison to the previous work, with standard deviation of 10 folds in brackets. } 
\label{tab:table2}

\begin{tabular}{|c|c|c|c|c|}
\hline
  & DS-1 & DS-2 & DS-3 & DS-4 \\
\hline
  Yesiler \cite{3.1Yes} & 0.897 (0.036) & \textbf{0.972 (0.021)} & 0.922 (0.0031) & 0.855 (0.097) \\
\hline
  Simple CNN & 0.907 (0.0290) & 0.851 (0.0726) & 0.935 (0.0386) & 0.858 (0.1078) \\
\hline
  Original RA based & \textbf{0.922 (0.0269)} & 0.941 (0.0152) & 0.909 (0.0274) & 0.742 (0.1036) \\
\hline
  Augmented RA based & 0.914 (0.0371) & 0.855 (0.0525) & \textbf{0.945 (0.0240)} & \textbf{0.927 (0.0988)} \\
\hline
\end{tabular}
\end{center}
\end{table*}

\begin{table*}[t]
\begin{center}
\caption{Average accuracy score values of the proposed methods and other designed experiments \protect\\in comparison to the previous work, with standard deviation of 10 folds in brackets. } 
\label{tab:table3}

\begin{tabular}{|c|c|c|c|c|c|}
\hline
  & DS-1 & DS-2 & DS-3 & DS-4 \\
\hline
  Yesiler \cite{3.1Yes} & 89.81\% (0.036) & \textbf{97.21\% (0.021)} & 92.29\% (0.031) & 85.83\% (0.100) \\
\hline
  Simple CNN & 90.79\% (0.0289) & 86.48\% (0.0606) & 93.50\% (0.0380) & 86.36\% (0.1009)  \\
\hline
  Original RA based & \textbf{92.25\% (0.0264)} & 94.10\% (0.0152) & 91.01\% (0.0270) & 75.52\% (0.0850) \\
\hline
  Augmented RA based & 91.52\% (0.0358) & 86.22\% (0.0498) & \textbf{94.58\% (0.0236)} & \textbf{92.92\% (0.0974)}  \\
\hline
  
\end{tabular}
\end{center}
\end{table*}

Inspired by the application of attention mechanism in fine-grained image recognition, this paper builds a network with reference to the Residual Attention Network proposed in \cite{2.4Wang} to capture subtle difference in four phonation modes. The attention module in Residual Attention Network consists of two parts: a mask branch and a trunk branch. The trunk branch performs feature processing while the mask branch is used for controlling gates of the trunk branch. The input data $x$ is sent into trunk branch and $T(x)$ is the output. After getting the weighted attention map $M(x)$, the values in which ranges within $[0, 1]$, an element-wise operation is performed with $T(x)$ produced by the trunk branch. The final output of the module is:

\begin{eqnarray}
H_{i,c}(x) &=& (1 + M_{i,c}(x)) * T_{i,c}(x)
\label{eq:eq1}
\end{eqnarray}

\noindent where $i$ ranges over all spatial positions and $c \in \{1, ..., C \}$ is the index of the channel.

The attention module enables the network focus on a specific area, and can also enhance its characteristics. The bottom-up top-down feedforward structure is used to unfold the feedforward and feedback attention process into a single feedforward process. The residual mechanism helps to mitigate the gradient vanishing problem.

Based on the idea mentioned above, we build a Residual Attention based network for automatic classification of phonation modes. The structure of the network can be seen in Figure \ref{fig:fig_network}. 

We first construct a convolutional neural network with frequency-biased filters. The network mainly consists of 4 convolutional layers and 2 fully connected layers. The sizes of the filters used in convolutional layers are shown in Figure \ref{fig:fig_network}. Due to the relatively small size of the spectrogram, we perform max pooling only after the first two convolutional layers to reduce the loss of valid data. Then, for the third and fourth convolutional layers, we add a soft mask branch similar to which in attention modules of the Residual Attention Network. 

Different from the networks for image classification, the proposed network is relatively shallow because of the limited training data. Moreover, for this task, the input data is spectrogram, each part of which makes sense because there is no noise or blank segment. Therefore, it is not necessary to stack attention modules to extract too many levels of attention information, and only one bottom-up top-down feedforward structure is adapted in the proposed network. 

\section{Experiments}

Three experiments are designed to verify the effects of the designed network and data augmentation, using the network without soft mask branch trained with augmented data (referred to as "Simple CNN"), and the Residual Attention based network trained with original data (referred to as "Original RA based") and augmented data (referred to as "Augmented RA based") respectively. The details of these experiments are described below, and the results are showed in Table \ref{tab:table2} and Table \ref{tab:table3}.

In this section, we first introduce the experimental setup, and then describe the three tasks and discuss experimental results.

\subsection{Experimental setup}
% Mel-spectrogram are extracted with LibROSA. 
The framework for the training process was developed in Python using PyTorch. Training data is divided as 64 samples in each mini-batch, and is trained with GPU Nvidia GTX1070. In order to make the training process more robust, an Adam optimizer is applied as an adaptive optimizer for better performance with weight decay rate of 0.0001. Cross entropy is used as the loss function. The learning rate is 0.001, and its annealing rate is set to 0.5 per 20 epochs. 

To compare with the work in \cite{3.1Yes}, ten-fold cross-validation is utilized to make the results more reliable. The training set is divided into ten subsets and in each fold, nine out of ten subsets are used for training, and the remained one subset  for  validation. After training, we use the test set to evaluate each model, and take the average value of ten folds as the experimental result.

\subsection{Experimental results and analysis}
% \subsection{Comparison of augmentation data and origin data}

\subsubsection{Validation of data augmentation}

By data augmentation in Section 3.2, the training set has been expanded to approximately twice. We verified the effect of data augmentation by training the designed network on original data (referred to as "Original RA based") and augmented data (referred to as "Augmented RA based") respectively.  

The experimental results show that the data augmentation is effective on relatively small datasets. For DS-4, the trained model has a growth of more than 15\% in classification accuracy. But on larger datasets, such as DS-1 and DS-3, there is no obvious effect. We infer that the augmented data does not bring in more useful information for these datasets. Besides, the accuracy is reduced on DS-2 by 8.61\%. By further exploration, the length of stable audio segments in DS-2 is shorter than the other three datasets. After the augmentation, some unstable and ambiguous samples are even introduced, which leads to the worse results on DS-2.

\begin{figure}[t]
\begin{minipage}[b]{1.0\linewidth}
  \centering
  \centerline{\epsfig{figure=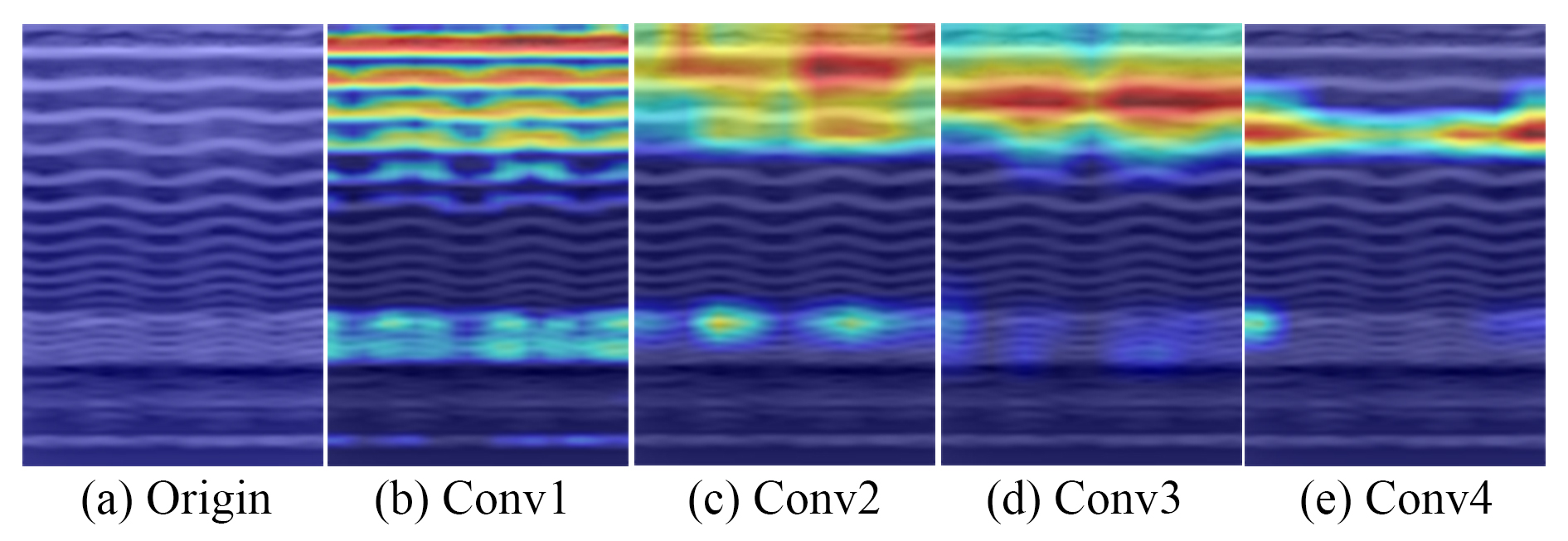,width=8.5cm}}
  \vspace{-0.15cm}
  \caption{Generated activation maps of flow mode after each convolutional layer in the network without soft mask branch.}
  \label{fig:fig_simplecnn}
\end{minipage}
\end{figure}

\begin{figure}[t]
\begin{minipage}[b]{1.0\linewidth}
  \centering
  \centerline{\epsfig{figure=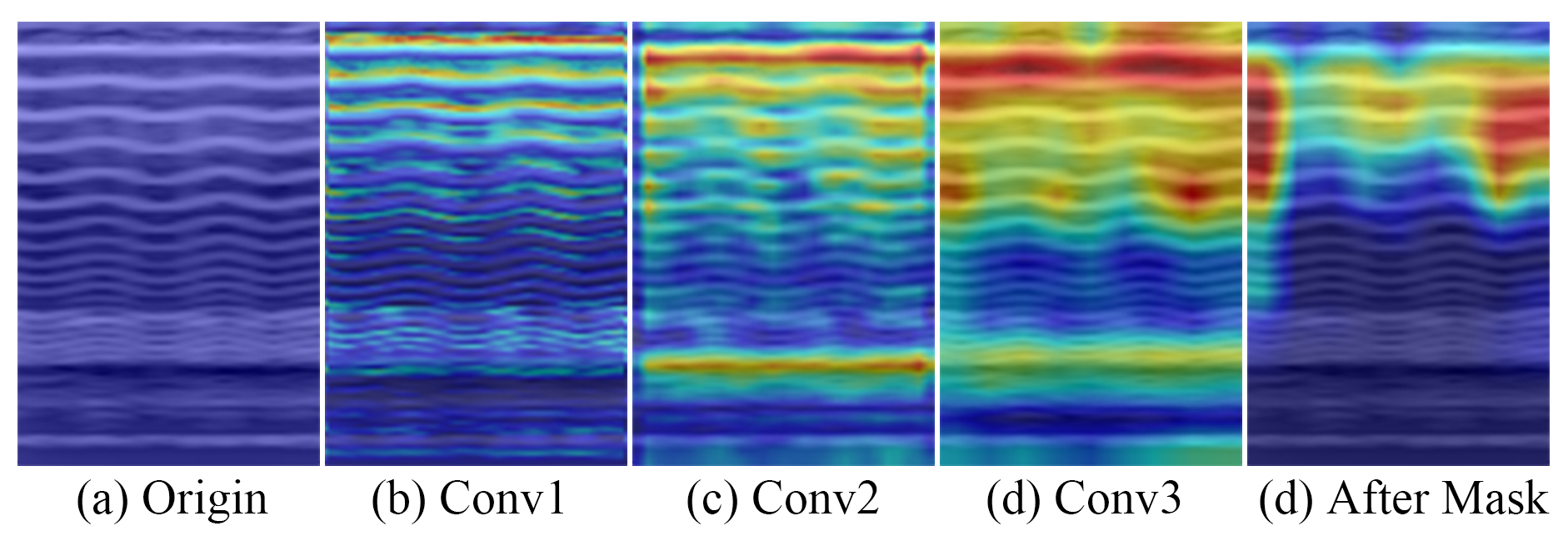,width=8.5cm}}
  \vspace{-0.15cm}
  \caption{Generated activation maps of flow mode after each convolutional layer in the Residual Attention based network.}
  \label{fig:fig_rabased}
\end{minipage}
\end{figure}

\subsubsection{Verification of the effect of the soft mask branch}

To validate the effect of the proposed attention mechanism on the models, we designed an experiment to compare the designed network with a similar network without relevant modules (referred to as "Simple CNN"). Because these modules are designed as a branch, namely soft mask branch, it is easy to remove them. For the remaining trunk network, the same super parameters are used to train models with the augmented datasets. 

The experimental results show that the accuracy of the models on DS-1, DS-3 and DS-4 improves to some extent by adopting the idea. Especially on DS-4, the F-measure of "Augmented RA based" is 0.069 higher than that of "Simple CNN". Even on DS-2, where the accuracy score is slightly worse, the F-measure achieves 0.855, 0.004 higher than that of "Simple CNN".

To further explain the use of the mask, in Section 4.3, we use Grad-CAM to analyze the gradient, visualizing the network attention of different layers. 

\subsubsection{Comparison with previous works}

We compare the results of our work with previous studies. Yesiler \cite{3.1Yes} trained a simple neural network model with hand-crafted features, and achieved relatively good results. Different from his work, the proposed method automatically learns the discriminating features from the time-frequency representation of the signal. 

The experimental results show that the trained models in three experiments surpass the baseline in most cases. It proves that the approach of extracting discriminative features through CNN automatically is effective. 

Among the results, there are 0.025, 0.023 and 0.072 F-measure improvement on DS-1, DS-3 and DS-4 respectively. For DS-2, the best result of the three experiments is "Original RA based", achieving 0.941 F-measure and 94.10\% accuracy, which is slightly worse than the previous work. For the most commonly used dataset DS-1, "Original RA based" achieves state-of-the-art result 0.922 for F-measure and 92.25\% for accuracy score) compared with previous attempts (0.897 F-measure and 89.81\% accuracy by Yesiler \cite{3.1Yes}, 75\% accuracy by Proutskova \cite{3.1prout}, 0.841 F-measure by Ioannidis and Rouas \cite{1.1rouas}, and 0.868 F-measure by Stoller and Dixon \cite{3.3stoller}). 

\subsection{Grad-CAM analysis}

Grad-CAM is a visualization technique for interpretability of deep convolutional networks proposed in 2016 \cite{3.4gradcam}. CNN are trained by back-propagation mechanism. Grad-CAM obtains the gradient value of the convolution kernel during the back-propagation, then multiplies the processed gradient value with the original feature map to get class activation map.

Taken flow mode as an example, we visualize the activation map during training with Grad-CAM method. As shown in Figure \ref{fig:fig_simplecnn}, for the model without soft mask branch, high attention is mainly focus on few frequency bands with significant fluctuations. On the contrary, the Residual Attention based model does decently well on Mel-spectrograms. From Figure \ref{fig:fig_rabased}, it can be seen that earlier gradient attention focuses on fundamental frequency and the harmonic parts of mid and high frequency band, while the latter focuses on the areas that more pertinent to characterize flow mode. The phenomenon is consistent with previous assumption, indicating that the regular fluctuations in mid and higher bands are dominant for discriminating the flow mode from other types.

\begin{figure}[h]
\begin{minipage}[b]{1.0\linewidth}
  \centering
  \centerline{\epsfig{figure=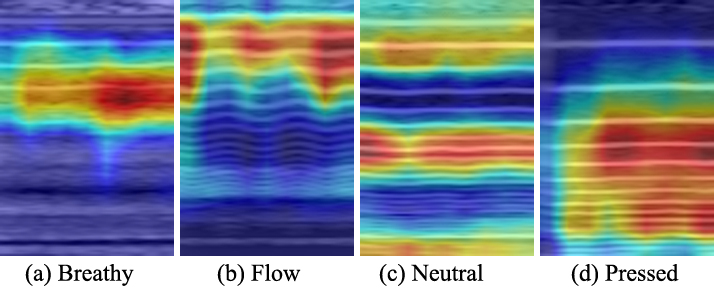,width=8.5cm}}
  \vspace{-0.12cm}
  \caption{Generated activation maps of four phonation modes with the Residual Attention based network.}
  \label{fig:fig_masked}
\end{minipage}
\end{figure}

Figure \ref{fig:fig_masked} shows the gradient attention maps of four phonation modes. (a) The attention of breathy mode mainly focuses on the irregular and foggy parts in high frequency. (b) The attention of flow mode mainly concentrates on the fundamental frequency and the harmonic parts with obvious regular vibration. (c) The attention of neutral mode evenly focuses on low, medium and high frequency, while (d) the attention of pressed mode mainly concentrates on the middle frequency areas where the energy is relatively concentrated.

\section{Conclusion}

In this study, we proposed a Residual Attention based network for automatic classification of phonation modes.The network consists of a simple convolutional network performing feature processing and a soft mask branch enabling the network focus on a specific area. Comparison experiments show the effectiveness of the proposed network, and the data augmentation is proved to be effective in some scenarios. Furthermore, visualization of the class activation map using Grad-CAM method demonstrates the enhancement behavior for dominant classification features in the task.

% \section{Acknowledgement}
% This work was supported by National Key R\&D Program of China (2019YFC1711800), NSFC (61671156).

\bibliographystyle{IEEEbib}
\bibliography{icme2020template}

\begin{thebibliography}{10}

\bibitem{1.1Sundberg}
J.~Sundberg and T.~D. Rossing,
\newblock {\em The science of singing voice},
\newblock ASA, 1990.

\bibitem{1.1Childers}
D.~G. Childers and C.~K. Lee,
\newblock ``Vocal quality factors: Analysis, synthesis, and perception,''
\newblock {\em The Journal of the Acoustical Society of America}, vol. 90, no.
  5, pp. 2394--2410, 1991.

\bibitem{3.1prout}
P.~Proutskova{,} C. Rhodes{,}~T. Crawford{,} and G.~Wiggins,
\newblock ``Breathy, resonant, pressed – automatic detection of phonation
  mode from audio recordings of singing,''
\newblock {\em Journal of New Music Research}, vol. 42, no. 2, pp. 171--186,
  2013.

\bibitem{1.1grillo}
E.~U. Grillo and K.~Verdolini,
\newblock ``Evidence for distinguishing pressed, normal, resonant, and breathy
  voice qualities by laryngeal resistance and vocal efficiency in vocally
  trained subjects,''
\newblock {\em Journal of Voice}, vol. 22, no. 5, pp. 546--552, 2008.

\bibitem{1.1millgaard}
M.~Millgård{,}~T. Fors{,} and J.~Sundberg,
\newblock ``Flow glottogram characteristics and perceived degree of phonatory
  pressedness,''
\newblock {\em Journal of Voice}, vol. 30, no. 3, pp. 287--292, 2016.

\bibitem{1.1airas}
M.~Airas and P.~Alku,
\newblock ``Comparison of multiple voice source parameters in different
  phonation types,''
\newblock in {\em Annual Conference of the International Speech Communication
  Association}, 2007.

\bibitem{1.1alku}
P.~Alku{,}~T. Bäckström{,} and E.~Vilkman,
\newblock ``Normalized amplitude quotient for parametrization of the glottal
  flow,''
\newblock {\em the Journal of the Acoustical Society of America}, vol. 112, no.
  2, pp. 701--710, 2002.

\bibitem{1.1sundberg2}
J.~Sundberg{,} M. Thalén{,}~P. Alku{,} and E.~Vilkman,
\newblock ``Estimating perceived phonatory pressedness in singing from flow
  glottograms,''
\newblock {\em Journal of Voice}, vol. 18, no. 1, pp. 56--62, 2004.

\bibitem{1.1kane1}
J.~Kane and C.~Gobl,
\newblock ``Identifying regions of non-modal phonation using features of the
  wavelet transform,''
\newblock in {\em Annual Conference of the International Speech Communication
  Association}, 2011.

\bibitem{1.1kane2}
J.~Kane and C.~Gobl,
\newblock ``Wavelet maxima dispersion for breathy to tense voice
  discrimination,''
\newblock {\em IEEE Transactions on Audio, Speech, and Language Processing},
  vol. 21, no. 6, pp. 1170--1179, 2013.

\bibitem{1.1hille}
J.~Hillenbrand{,} R.~A. Cleveland{,} and R.~L. Erickson,
\newblock ``Acoustic correlates of breathy vocal quality,''
\newblock {\em Journal of Speech, Language, and Hearing Research}, vol. 37, no.
  4, pp. 769--778, 1994.

\bibitem{1.1kadiri1}
S.~R. Kadiri and B.~Yegnanarayana,
\newblock ``Analysis and detection of phonation modes in singing voice using
  excitation source features and single frequency filtering cepstral
  coefficients ({SFFCC}),''
\newblock in {\em Interspeech}, 2018.

\bibitem{1.1kadiri2}
S.~R. Kadiri and B.~Yegnanarayana,
\newblock ``Breathy to tense voice discrimination using zero-time windowing
  cepstral coefficients ({ZTWCC}s),''
\newblock in {\em Interspeech}, 2018.

\bibitem{3.1rouas}
J.-L. Rouas and L.~Ioannidis,
\newblock ``Automatic classification of phonation modes in singing voice:
  towards singing style characterisation and application to ethnomusicological
  recordings,''
\newblock in {\em Interspeech}, 2016.

\bibitem{3.1Yes}
F.~Yesiler,
\newblock ``Analysis and automatic classification of phonation modes in
  singing,''
\newblock M.S. thesis, Universitat Pompeu Fabra, 2018.

\bibitem{2.3Pons}
J.~Pons{,}~T. Lidy{,} and X.~Serra,
\newblock ``Experimenting with musically motivated convolutional neural
  networks,''
\newblock in {\em International Workshop on Content-Based Multimedia Indexing},
  2016.

\bibitem{2.4Wang}
F.~Wang{,} M. Jiang{,} C. Qian{,} S. Yang{,} C. Li{,} H. Zhang{,}~X. Wang{,}
  and X.~Tang,
\newblock ``Residual attention network for image classification,''
\newblock in {\em IEEE Conference on Computer Vision and Pattern Recognition},
  2017.

\bibitem{1.1rouas}
J.-L. Rouas and L.~Ioannidis,
\newblock ``Automatic classification of phonation modes in singing voice:
  Towards singing style characterisation and application to ethnomusicological
  recordings,''
\newblock in {\em Interspeech}, 2016.

\bibitem{3.3stoller}
D.~Stoller and S.~Dixon,
\newblock ``Analysis and classification of phonation modes in singing,''
\newblock in {\em International Society for Music Information Retrieval
  Conference}, 2016.

\bibitem{3.4gradcam}
R.~R. Selvaraju{,} M. Cogswell{,} A. Das{,} R. Vedantam{,}~D. Parikh{,} and
  D.~Batra,
\newblock ``Grad-cam: Visual explanations from deep networks via gradient-based
  localization,''
\newblock in {\em IEEE International Conference on Computer Vision}, 2017.

\end{thebibliography}

\end{document}